\documentclass[conference]{IEEEtran}
\IEEEoverridecommandlockouts
\usepackage[comma, sort&compress,square,numbers]{natbib}
\usepackage{amsmath,amssymb,amsfonts}
\usepackage{algorithmic}
\usepackage{graphicx}
\usepackage{textcomp}
\usepackage{booktabs}
\usepackage{multirow}
\usepackage{xcolor}

\usepackage{lipsum}
\usepackage{flushend}
\usepackage{tabularx}
\usepackage{hyperref}
\usepackage[linesnumbered,ruled,lined,commentsnumbered]{algorithm2e}
\usepackage{color,soul}
\usepackage[T1]{fontenc}
\usepackage[utf8]{inputenc}

\usepackage[small]{caption}
\usepackage{url}
\usepackage[inline]{enumitem}
\usepackage[inkscapelatex=false]{svg}
\usepackage{float}
\def\BibTeX{{\rm B\kern-.05em{\sc i\kern-.025em b}\kern-.08em
    T\kern-.1667em\lower.7ex\hbox{E}\kern-.125emX}}

\begin{document}

\title{Compressing Deep Image Super-resolution Models\\
\thanks{The authors appreciate the funding from Netflix Inc., University of Bristol, and the UKRI MyWorld Strength in Places Programme (SIPF00006/1).}
}


\author{\IEEEauthorblockN{Yuxuan Jiang,
Jakub Nawała,
Fan Zhang, and 
David Bull}
\IEEEauthorblockA{\textit{Visual Information Laboratory,
University of Bristol, Bristol, BS1 5DD, UK}\\
\textit {\{yuxuan.jiang, jakub.nawala, fan.zhang, dave.bull\}@bristol.ac.uk}}
}

\maketitle

\begin{abstract}
Deep learning techniques have been applied in the context of image super-resolution (SR), achieving remarkable advances in terms of reconstruction performance. Existing techniques typically employ highly complex model structures which result in large model sizes and slow inference speeds. This often leads to high energy consumption and restricts their adoption for practical applications. To address this issue, this work employs a three-stage workflow for compressing deep SR models which significantly reduces their memory requirement. Restoration performance has been maintained through teacher-student knowledge distillation using a newly designed distillation loss. We have applied this approach to two popular image super-resolution networks, SwinIR and EDSR, to demonstrate its effectiveness. The resulting compact models, SwinIRmini and EDSRmini, attain an 89\% and 96\% reduction in both model size and floating-point operations (FLOPs) respectively, compared to their original versions. They also retain competitive super-resolution performance compared to their original models and other commonly used SR approaches. The source code and pre-trained models for these two lightweight SR approaches are released at \url{https://pikapi22.github.io/CDISM/}.
\end{abstract}


\section{Introduction}

Image super-resolution (SR) has attracted growing research interest over the past few decades. It represents the task of generating a high spatial resolution image from a low-resolution version, with the aim of reconstructing with optimal perceptual quality, accurately recovering spatial detail. It has been widely employed in various image and video processing applications including medical imaging, image restoration and enhancement, and picture coding \cite{wang2020deep, bull2021intelligent, afonso2018video, zhang2021vistra2}. SR can be conventionally achieved by using various linear and non-linear filters \cite{yang2010image, schulter2015fast,afonso2017low}, while learning-based SR has become more popular recently due to its superior reconstruction performance.
\par
Learning-based SR algorithms can be divided into two major categories: CNN-based \cite{dong2015image, kim2016accurate, lim2017enhanced, zhang2018image, ma2020cvegan, ma2020mfrnet} and Transformer-based approaches \cite{liang2022vrt, wang2022uformer, liang2021swinir, conde2022swin2sr}. The former commonly leverages a Convolutional Neural Network (CNN), which typically comprises a number of consecutive convolutional layers connected with activation functions. Typical examples include SRCNN \cite{dong2015image}, VDSR \cite{kim2016accurate}, EDSR \cite{lim2017enhanced} and RT4KSR \cite{zamfir2023towards}. More recently, with the invention of Vision Transformer (ViT) networks \cite{vaswani2017attention} which exploit self-attention mechanisms to capture more context interaction information, reconstruction performance has been further improved by integrating ViT into the SR framework (notable contributions include SwinIR \cite{liang2021swinir} and Swin2SR \cite{conde2022swin2sr}).

\begin{figure}[t]
    \centering
    \includegraphics[width=1\linewidth]{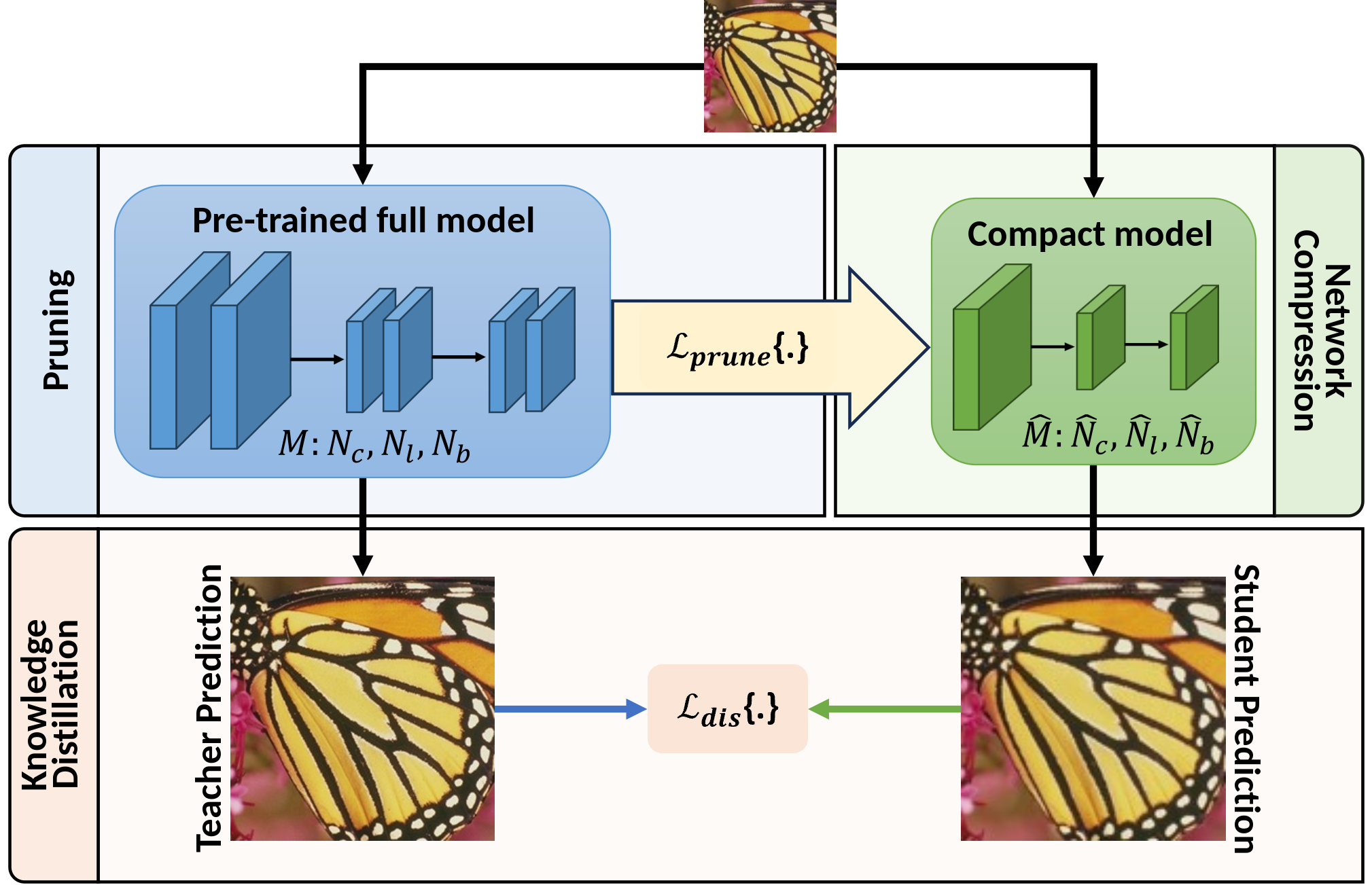}
    \caption{\small The proposed workflow for compressing SR networks. The pruning process obtains the corresponding compact model $\hat{M}$ from its intricate counterpart $M$. Here $N_c$, $N_l$ and $N_b$ represent the channel number, the layer number and the block number, respectively. $\hat{N}_c$, $\hat{N}_l$ and $\hat{N}_b$ denote the same but for the model after compression. $\mathcal{L}_{prune}\{.\}$ and $\mathcal{L}_{dis}\{.\}$ are the loss functions used in pruning and knowledge distillation processes. 
    }
    \label{fig2}
\end{figure}
Although the aforementioned learning-based SR algorithms have significantly improved performance compared to conventional filters, the former is often associated with high computational complexity, leading to slow run-time and large memory requirements. For example, one of the best performers, SwinIR, requires approximately 11.8M parameters \cite{liang2021swinir}, while the widely used EDSR approach is based on a model with 43M parameters \cite{lim2017enhanced}. These highly complex models demand significant computational resources during their development and inference, restricting their adoption in practical applications. Hence, there is a pressing necessity to perform complexity reduction for these approaches while maintaining their excellent SR performance.
\par
In recent reports, complexity reduction has been performed through sophisticated manual modifications of network architectures \cite{zamfir2023towards, guo2023asconvsr, chu2021fast}. It has also been achieved using model compression techniques \cite{cheng2017survey, kirchhoffer2021overview} such as model pruning \cite{zhu2017prune, kong2022residual} and knowledge distillation (KD) \cite{hinton2015distilling}. Pruning approaches eliminate redundancies within the original model and compress it into a more compact form, resulting in a substantial reduction in model size. KD, on the other hand, involves transferring knowledge from a large teacher model (e.g., the original version) to a smaller student model (e.g., the compact one). These techniques have been proven to be effective when adopted separately for various image processing applications, such as image classification \cite{chen2021distilling, jin2023multi, beyer2022knowledge}, image restoration \cite{fang2022cross} and video compression \cite{kwan2023hinerv}. However, only a few works (mainly for video frame interpolation \cite{ding2021cdfi, morris2023st}) have investigated the combination of model pruning and knowledge distillation. 

In this context, with the focus on image super-resolution, we propose a new network compression framework, illustrated in Fig. \ref{fig2}, which integrates both model pruning and knowledge distillation to significantly reduce the model complexity while maintaining SR performance. This approach first applies sparsity inducing optimisation to the original network before compressing it into a compact model based on a new parameter distribution analysis method. The performance of the compressed model is then further improved through knowledge distillation with a modified loss. To the best of our knowledge, this is the first attempt that combines both pruning and distillation techniques for SR model compression. We have applied this to two popular SR models, EDSR and SwinIR, and their resulting compact models achieve significant model size and FLOPs (up to 96\%) reductions,  significantly outperforming other SR methods with similar complexity figures.

\section{Proposed Algorithm}
\label{sec:method}

\begin{figure}[t]
    \centering
    \includegraphics[width=1\linewidth]{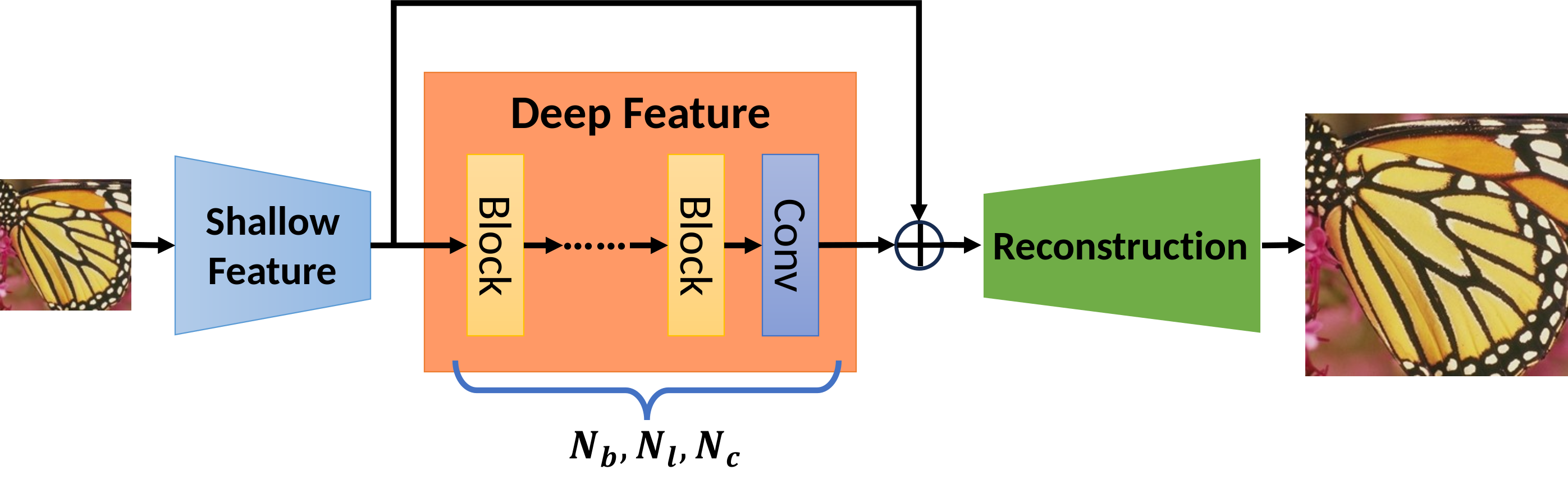}
    \caption{\small The basic blueprint of a modern image SR network.}
    \label{fig1}
\end{figure}

Fig. \ref{fig1} illustrates a commonly employed high-level network architecture for image super-resolution. It comprises three integral components: a shallow feature extraction module, a deep feature extraction module, and an image reconstruction module. The shallow feature extraction module usually employs a small number of convolutional layers to extract shallow features containing essential low-frequency information. The core of the network lies within the deep feature extraction module, which meticulously obtains intricate and high-level features, collectively holding a pivotal role in shaping the system's overall performance and capabilities. Ultimately, both shallow and deep features converge within the reconstruction module to facilitate the creation of high-quality image reconstructions. Since the network structure (mainly in the deep feature extraction module) consists of a stack of basic processing blocks, its complexity and performance are intimately tied to the number of channels $N_c$, the layer counts (excluding the convolutional layer before the output) $N_l$ within each block, and the total number of such blocks $N_b$. 

The proposed complexity reduction workflow for deep SR models is shown in Fig. \ref{fig2}, which consists of three primary stages: model pruning, network compression, and knowledge distillation. The algorithm underpinning this is detailed in the following subsections.   

\subsection{Model Pruning}

In order to obtain a condensed version of a model, we start with the original pre-trained model, and fine-tune it using the following loss function,
\begin{equation}
    \mathcal{L}_{prune} = \sqrt{(I_\mathit{SR}-I_{gt})^2+\epsilon^2}+\lambda \|\theta\|_{1},
    \label{eq1}
\end{equation}
where $I_{gt}$ represents the ground-truth target image, $I_\mathit{SR}$ corresponds to the super-resolved output, $\epsilon$ is set to $10^{-3}$, and $\lambda > 0$ is the regularisation constant, set to $10^{-4}$ following \cite{morris2023st}. The parameters of the initial model are represented by $\theta$, and $\|\cdot\|_{1}$ refers to the $L1$ norm regularisation term, which serves to promote network sparsity, as discussed in \cite{chen2021orthant}. Such sparsity information can be used as a guide to removing redundant layers in the network. We adopt the OBProx-SG \cite{chen2021orthant} solver to perform the optimisation. Eventually, a density ratio $d$, the ratio of non-zero parameters, is obtained. In contrast to the approach taken in \cite{ding2021cdfi, morris2023st}, where they computed the density ratio for each layer, we calculate $d$ for the whole deep feature module. We then use this ratio in the next step of complexity reduction.

\subsection{Network Compression}

Contrary to the method described in \cite{morris2023st}, where the compression only focuses on the channel level, here we further analyse three hyperparameters $N_c$ (the number of channels), $N_l$ (the number of layers), and $N_b$ (the number of deep feature extraction blocks). Specifically, we consider the total number of model parameters for the deep feature extraction module, $P_\mathit{DF}$, which can be approximately\footnote{If the block structure is unique, it needs to be analysed on a case-by-case basis.} written as:
\begin{equation}
     P_\mathit{DF} \approx k{N}_b ({N}_l + 1){N}^2_c.
     \label{eq:Ptotal}
\end{equation}
Here, $k$ is an approximate constant for a specific model structure. With the density ratio $d$ calculated during the model pruning stage, we can obtain a compact model by updating these three key hyperparameters to satisfy the following: 
\begin{equation}
    \frac{\hat{P}_\mathit{DF}}{P_\mathit{DF}} \approx d,
    \label{eq:ratio}
\end{equation}
in which $\hat{P}_\mathit{DF}$ represents the total number of parameters of the compact deep feature extraction module, 
\begin{equation}
     \hat{P}_\mathit{DF} \approx k \hat{N}_b (\hat{N}_l + 1)\hat{N}^2_c.
     \label{eq:Ptotal1}
\end{equation}

If we substitute eq. (\ref{eq:Ptotal}) and (\ref{eq:Ptotal1}) into eq. (\ref{eq:ratio}), we get 

\begin{equation}
    \frac{\hat{N}_b (\hat{N}_l + 1)\hat{N}^2_c}{{N}_b ({N}_l + 1){N}^2_c} \approx d.
    \label{eq:ratio1}
\end{equation}
It is noted that the influence of these three hyperparameters on the model performance and size is not identical---the reduction of $N_b$ and $N_l$ leads to a greater model performance reduction, while $N_c$ has a greater influence on model size. Based on these observations, we achieve model compression by empirically assigning
\begin{equation}
         \frac{\hat{N}_b}{N_b} \approx \sqrt[6]{d} \text{ and } 
         \frac{\hat{N}_l + 1}{{N}_l + 1} \approx \sqrt[6]{d}, \quad \hat{N}_b,  \hat{N}_l \in \mathbb{N}^+
         \label{eq:NbNl}
\end{equation}
After obtaining $\hat{N}_b$ and $\hat{N}_l$, $\hat{N}_c$ can be derived from
\begin{equation}
        \frac{\hat{N}_c}{N_c} \approx  \
         \sqrt[3]{d \frac{{N}_b ({N}_l + 1)}{\hat{N}_b (\hat{N}_l + 1)}}, \quad \hat{N}_c \in \mathbb{N}^+
           \label{eq:Nc}
\end{equation}
After that, we accordingly adjust the shallow feature extractor and reconstruction module based on $\hat{N}_c$  to maintain the network integrity.

\subsection{Knowledge Distillation}

After obtaining the compact model, using a knowledge distillation approach similar to \cite{morris2023st}, we further improve the pruned model's performance by employing the pre-trained original model as a ``teacher'' to instruct the training process. Specifically, the total loss $\mathcal{L}_{total}$ for knowledge distillation is given as follows,
\begin{equation}
    \mathcal{L}_{total} = \alpha\mathcal{L}_{stu}(I_{stu}, I_{gt}) + \mathcal{L}_{dis}(I_{stu}, I_{tea}),
    \label{Ltotal}
\end{equation}
where $\mathcal{L}_{stu}$ denotes the original loss between the ground truth $I_{gt}$ and the student model’s prediction $I_{stu}$ (which was used for training the original full SR model), $\alpha$ is a tunable weight, and $\mathcal{L}_{dis}$ stands for the loss between the student $I_{stu}$ and the teacher’s predictions $I_{tea}$. In this work, the distillation loss, $\mathcal{L}_{dis}$, is calculated as below, which is inspired by \cite{morris2023st,conde2022swin2sr,niklaus2018context}:
\begin{eqnarray}
    \mathcal{L}_{dis}(I_{stu}, I_{tea}) &=& \mathcal{L}_{Lap}(I_{stu}, I_{tea})\nonumber \\
    & &+\; \mathcal{L}_{Lap}(\mathit{HF}(I_{stu}), \mathit{HF}(I_{tea})),
    \label{Ldis}
\end{eqnarray}
where $\mathcal{L}_{Lap}$ is the Laplacian loss \cite{niklaus2018context} and $\mathit{HF}(\cdot)$ represents the high-frequency features extracted by a 5$\times$5 Gaussian blur kernel function. The high-frequency feature loss is included to further enhance the sharpness and overall quality of the output.

\begin{figure*}[htbp]
    \captionsetup{justification=centering}
	\centering
	\begin{minipage}{0.15\linewidth}
		\centering
		\includegraphics[width=0.9\linewidth]{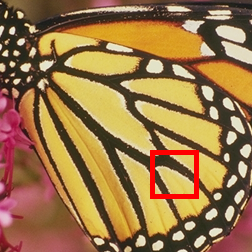}
            \caption*{\small butterfly (x2)}
	\end{minipage}
	\begin{minipage}{0.15\linewidth}
		\centering
		\includegraphics[width=0.9\linewidth]{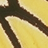}
            \caption*{\small GT}

	\end{minipage}
	\begin{minipage}{0.15\linewidth}
		\centering
		\includegraphics[width=0.9\linewidth]{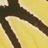}
            \caption*{\small EDSR\_baseline}
	\end{minipage}
 	\begin{minipage}{0.15\linewidth}
		\centering
		\includegraphics[width=0.9\linewidth]{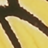}
            \caption*{\small \textbf{EDSRmini} (ours)}
	\end{minipage}
 	\begin{minipage}{0.15\linewidth}
		\centering
		\includegraphics[width=0.9\linewidth]{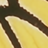}
            \caption*{\small SwinIR\_LW}
	\end{minipage}
  	\begin{minipage}{0.15\linewidth}
		\centering
		\includegraphics[width=0.9\linewidth]{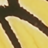}
            \caption*{\small \textbf{SwinIRmini} (ours)}
	\end{minipage}
	
	\begin{minipage}{0.15\linewidth}
		\centering
		\includegraphics[width=0.9\linewidth]{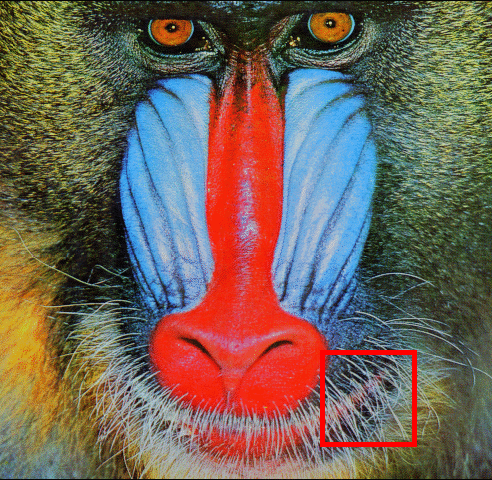}
            \caption*{\small baboon (x2)}
	\end{minipage}
	\begin{minipage}{0.15\linewidth}
		\centering
		\includegraphics[width=0.9\linewidth]{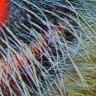}
            \caption*{\small GT}
	\end{minipage}
        \begin{minipage}{0.15\linewidth}
		\centering
		\includegraphics[width=0.9\linewidth]{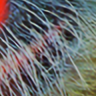}
            \caption*{\small EDSR\_baseline}
	\end{minipage}
        \begin{minipage}{0.15\linewidth}
		\centering
		\includegraphics[width=0.9\linewidth]{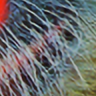}
            \caption*{\small \textbf{EDSRmini} (ours)}
	\end{minipage}
 	\begin{minipage}{0.15\linewidth}
		\centering
		\includegraphics[width=0.9\linewidth]{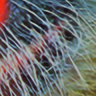}
            \caption*{\small SwinIR\_LW}
	\end{minipage}
  	\begin{minipage}{0.15\linewidth}
		\centering
		\includegraphics[width=0.9\linewidth]{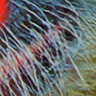}
            \caption*{\small \textbf{SwinIRmini} (ours)}
	\end{minipage}
    \caption{\small Example of visual comparison between our compact models and their original counterparts.}
    \label{Qualitative}
    \vspace{-10pt}
\end{figure*}

\section{Results and Discussion}
\label{sec:results}

In order to demonstrate the effectiveness of our complexity reduction workflow, we applied it to two popular image super-resolution models: EDSR  \cite{lim2017enhanced} and SwinIR \cite{liang2021swinir}. The former is a widely used CNN-based model, while SwinIR is Transformer-based and offers the state-of-the-art SR performance. The compact models are obtained from their existing lightweight versions, \textit{EDSR\_baseline} and \textit{SwinIR\_lightweight (LW)}, as reported by their original authors. We refer to our compact models as \textit{EDSRmini} and \textit{SwinIRmini}, respectively.

\subsection{Experimental Setup} 

We use the same training set, DIV2K \cite{agustsson2017ntire}, for both model pruning and knowledge distillation, as in \cite{lim2017enhanced,liang2021swinir}. Specifically, 
similarly to \cite{ding2021cdfi}, 100 images from the DIV2K \cite{agustsson2017ntire} dataset are used to conduct model pruning, and the whole DIV2K is employed in the knowledge distillation step. During training,  we use the AdaMax optimizer with $\beta_{1} = 0.9$ and $\beta_{2} = 0.99$. The hyperparameter $\alpha$ in eq. (\ref{Ltotal}) is set to 0.1, following \cite{morris2023st}. 

For compressing the SwinIR model, by pruning the SwinIR\_LW network for 100 epochs, a density of approximately 0.089 is obtained and considered as the compression rate. Based on this, $\hat{N}_c$, $\hat{N}_l$, and $\hat{N}_b$ for SwinIRmini are calculated according to eq. (\ref{eq:NbNl}) and (\ref{eq:Nc}), and their values are 24, 4, and 3, respectively. The resulting total number of parameters for the SwinIRmini is 98.8K (878K for SwinIR\_LW). Similarly, by optimising the EDSR\_baseline network, a density of approximately 0.03 is achieved, and  $\hat{N}_c$, $\hat{N}_l$ and $\hat{N}_b$ for EDSRmini are calculated as 16, 1, and 8, respectively, with 49.6K parameters in total (1.37M for EDSR\_baseline). 

The evaluation is performed on Set5 \cite{bevilacqua2012low} and Set14 \cite{zeyde2012single} databases, which are commonly employed to benchmark super-resolution models. Two widely used quality metrics, peak signal-to-noise ratio (PSNR) and structural similarity (SSIM) \cite{wang2004image}, are used to measure the model performance. All experiments are conducted using an NVIDIA RTX 3090 GPU.

\begin{table}[t]
\scriptsize
\centering
\caption{Performance comparison between our approaches and several other methods on two benchmark datasets. PSNR/SSIM values on the Y channel are reported on each dataset. \#Ps and FLOPs stand for the total number of network parameters and floating-point operations, respectively. FLOPs are measured under the setting of upscaling SR images to 1280$\times$720 resolution on the $\times$2 scale.}
\begin{tabular}{r|c|c|cc}
\toprule
\multicolumn{1}{r|}{\multirow{2.5}{*}{\textbf{Model}}} & \multicolumn{1}{c|}{\multirow{2.5}{*}{\textbf{\#Ps(M)}}} & \multicolumn{1}{c|}{\multirow{2.5}{*}{\textbf{FLOPs(G)}}} & \multicolumn{1}{c|}{\textbf{Set5}} & \multicolumn{1}{c}{\textbf{Set14}} \\ 
 &  &  & \multicolumn{1}{c|}{\textbf{PSNR\big\uparrow/SSIM\big\uparrow}} & \multicolumn{1}{c}{\textbf{PSNR/SSIM}} \\ \midrule
{SRCNN \cite{dong2015image}} & \multicolumn{1}{c|}{0.057} & \multicolumn{1}{c|}{13.2} & \multicolumn{1}{c|}{36.66/0.9542} & 32.45/0.9067 \\ \midrule
{LapSRN \cite{lai2017deep}} & \multicolumn{1}{c|}{0.251} & \multicolumn{1}{c|}{29.9} & \multicolumn{1}{c|}{37.52/0.9590} & 32.99/0.9124 \\ \midrule
{CARN \cite{ahn2018fast}} & \multicolumn{1}{c|}{1.59} & \multicolumn{1}{c|}{222.8} & \multicolumn{1}{c|}{37.76/0.9590} & 33.52/0.9166 \\ \midrule
{LatticeNet \cite{luo2020latticenet}} & \multicolumn{1}{c|}{0.756} & \multicolumn{1}{c|}{169.5} & \multicolumn{1}{c|}{38.15/0.9610} & 33.78/0.9193 \\ \midrule
{IMDN \cite{hui2019lightweight}} & \multicolumn{1}{c|}{0.694} & \multicolumn{1}{c|}{158.8} & \multicolumn{1}{c|}{38.00/0.9605} & 33.63/0.9177 \\ \midrule
{RLFN-S \cite{kong2022residual}} & \multicolumn{1}{c|}{0.454} & \multicolumn{1}{c|}{68.1} & \multicolumn{1}{c|}{38.05/0.9607} & 33.68/0.9172 \\ \midrule
{FALSR-B \cite{chu2021fast}} & \multicolumn{1}{c|}{0.326} & \multicolumn{1}{c|}{74.7} & \multicolumn{1}{c|}{37.61/0.9585} & 33.29/0.9143 \\ \midrule
{RT4KSR-XL \cite{zamfir2023towards}} & \multicolumn{1}{c|}{0.092} & \multicolumn{1}{c|}{n/a} & \multicolumn{1}{c|}{36.83/0.9545} & 33.46/0.9197\\ \midrule
{EDSR \cite{lim2017enhanced}} & \multicolumn{1}{c|}{43} & \multicolumn{1}{c|}{9387.0} & \multicolumn{1}{c|}{38.11/0.9601} & 33.92/0.9195\\ \midrule
{EDSR\_baseline \cite{lim2017enhanced}} & \multicolumn{1}{c|}{1.37} & \multicolumn{1}{c|}{316.3} & \multicolumn{1}{c|}{37.99/0.9604} & 33.57/0.9175\\ \midrule
{\textbf{EDSRmini (ours)}} & \multicolumn{1}{c|}{\textbf{0.049}} & \multicolumn{1}{c|}{\textbf{11.7}} & \multicolumn{1}{c|}{\textbf{37.67/0.9597}} & \textbf{33.21/0.9141} \\ \midrule
{SwinIR \cite{liang2021swinir}} & \multicolumn{1}{c|}{11.8} & \multicolumn{1}{c|}{2301.0} & \multicolumn{1}{c|}{38.35/0.9620} & 34.14/0.9227 \\ \midrule
{SwinIR\_LW \cite{liang2021swinir}} & \multicolumn{1}{c|}{0.878} & \multicolumn{1}{c|}{195.6} & \multicolumn{1}{c|}{38.14/0.9611} & 33.86/0.9206 \\ \midrule
{\textbf{SwinIRmini (ours)}} & \multicolumn{1}{c|}{\textbf{0.099}} & \multicolumn{1}{c|}{\textbf{22.0}} & \multicolumn{1}{c|}{\textbf{37.88/0.9601}} & \textbf{33.60/0.9187} \\ \bottomrule
\end{tabular}
\label{tbl1}
\end{table}

\subsection{Quantitative Evaluation}

\begin{figure}[htbp]
    \centering
    \includegraphics[width=1\linewidth]{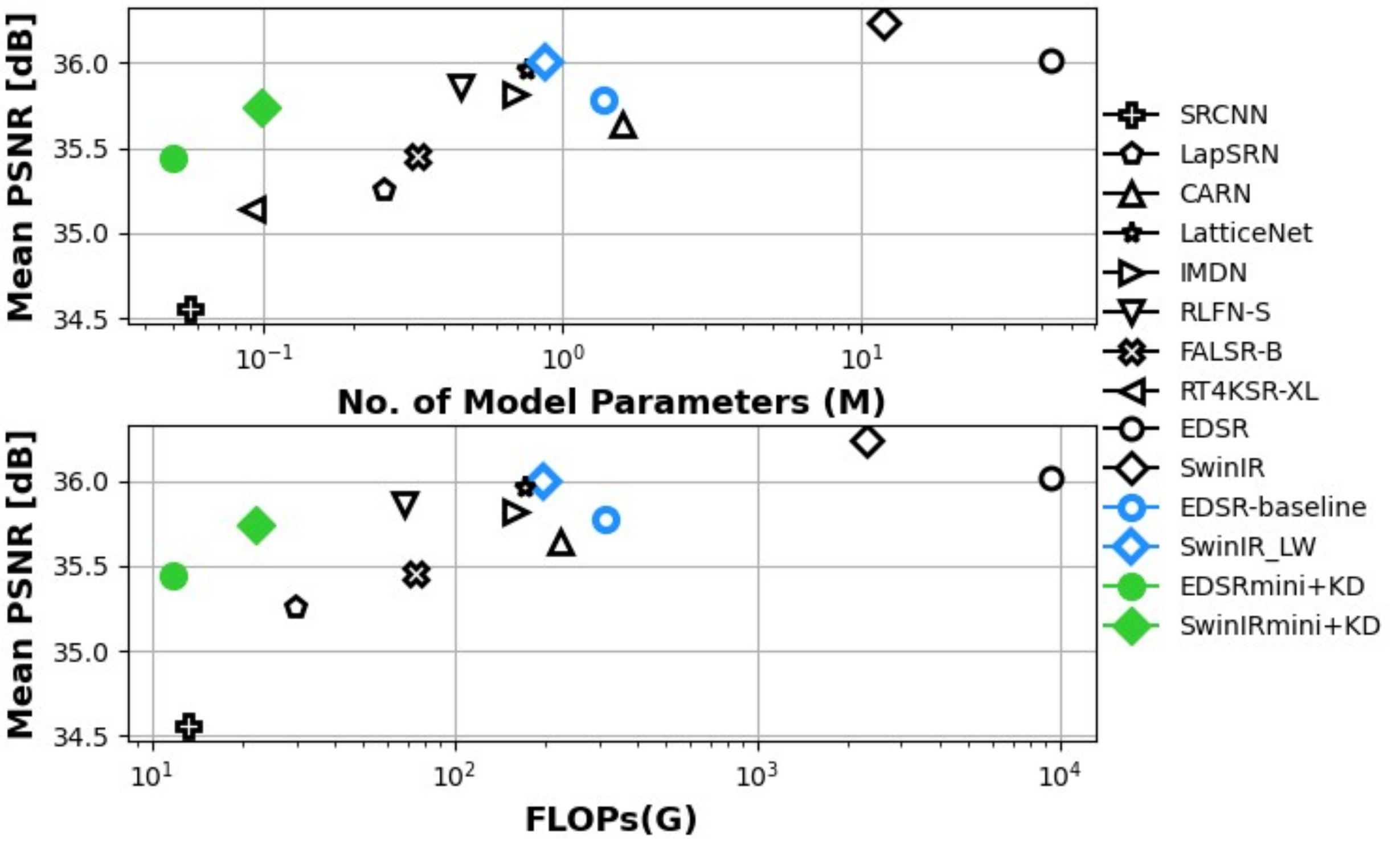}
    \caption{\small (Top) Average PSNR scores on Set5 and Set14 for the models presented in Table \ref{tbl1} versus their corresponding  numbers of model parameters. Results for our two compact models are marked with solid symbols. (Bottom) The plot between the performance and FLOPs based on Set5 and Set14.}
    \label{fig3}
\end{figure}

The results of the quantitative comparison between our approach and a number of existing deep SR approaches are summarised in TABLE \ref{tbl1}. It is noted that the performance results and complexity figures for all the benchmark results are taken from their original publications. It can be observed that the resulting compact models, EDSRmini and SwinIRmini, offer much smaller model sizes, 4\% and 11\% compared to their baseline models, EDSR\_baseline and SwinIR\_LW, respectively. They also require fewer FLOPs, 4\% and 11\% of their baselines. However, the average performance losses are minimal, 0.34 dB and 0.26 dB across both databases for EDSRmini and SwinIRmini, compared to their original models. Fig. \ref{fig3} provides a more intuitive illustration, plotting the average PSNR values of all the benchmarked SR models against the number of model parameters and FLOPs. It can be observed from both subfigures that the EDSRmini and SwinIRmini achieve a superior trade-off between complexity and performance. For example, SwinIRmini outperforms RT4KSR-XL with a comparable number of parameters by nearly 0.6 dB. 

\par

\subsection{Qualitative Evaluation}

Examples of SR outputs using both our compact models and their corresponding original versions are shown in Fig. \ref{Qualitative} for visual comparison. The results generated by both the compact models and their counterparts are largely indistinguishable, demonstrating the effectiveness of our complexity reduction approach. The proposed method not only significantly reduces the number of parameters and FLOPs, but also preserves the excellent interpolation performance of the original models.

\section{Conclusion}

\label{sec:conclusion}

This paper presents a new workflow for compressing image super-resolution (SR) models based on model compression and knowledge distillation. The proposed approach has been applied to two different SR approaches, EDSR (CNN-based) and SwinIR (Transformer-based), achieving consistent and significant complexity reduction results: up to 96\% in terms of model size and FLOPs. At the same time, the competitive performance of their original models has been maintained alongside the reduced complexity, which demonstrates the effectiveness of the proposed workflow. Future endeavours will extend this work to other low level computer vision tasks.

\small

\bibliographystyle{ieeetr}
\bibliography{refs}

\end{document}